\documentclass[prl,superscriptaddress,amsmath,amssymby,longbibliography,reprint]{revtex4-2}

\usepackage[english]{babel}
\PassOptionsToPackage{english}{babel}
\usepackage[utf8]{inputenc}
\usepackage{color}
\usepackage{amssymb,graphicx,xcolor,units,hyperref}
\usepackage{booktabs}
\usepackage{soul}
\usepackage{float}
\usepackage{siunitx}

\usepackage{hyperref}
\usepackage{makecell}

\usepackage{caption}
\hypersetup{
    bookmarks=false,         
    unicode=false,          
    pdftoolbar=false,        
    pdfmenubar=true,        
    pdffitwindow=false,     
    pdfstartview={FitH},    
    pdftitle={Superconductivity from a melted insulator},    
    pdfsubject={},   
    pdfcreator={},   
    pdfproducer={}, 
    pdfkeywords={}, 
    pdfnewwindow=true,      
    colorlinks=true,       
    linkcolor=black,          
    citecolor=blue,        
    filecolor=magenta,      
    urlcolor=blue           
}

\begin{document}
\selectlanguage{english} 

\title{Superconductivity from a melted insulator} 

\author{S. Mukhopadhyay}
\thanks{Equal contribution}
\affiliation{IST Austria, Am Campus 1, 3400 Klosterneuburg, Austria}
\author{J. Senior}
\thanks{Equal contribution}
\affiliation{IST Austria, Am Campus 1, 3400 Klosterneuburg, Austria}
\author{J. Saez-Mollejo}
\affiliation{IST Austria, Am Campus 1, 3400 Klosterneuburg, Austria}
\author{D. Puglia}
\affiliation{IST Austria, Am Campus 1, 3400 Klosterneuburg, Austria}
\author{M. Zemlicka}
\affiliation{IST Austria, Am Campus 1, 3400 Klosterneuburg, Austria}
\author{J. Fink}
\affiliation{IST Austria, Am Campus 1, 3400 Klosterneuburg, Austria}
\author{A.P. Higginbotham}
\email{andrew.higginbotham@ist.ac.at}
\affiliation{IST Austria, Am Campus 1, 3400 Klosterneuburg, Austria}


\maketitle 

\textbf{Quantum phase transitions typically result in a broadened critical or crossover region at nonzero temperature \cite{sachdev_quantum_2011}.
Josephson arrays are a model of this phenomenon \cite{sondhi_continuous_1997}, exhibiting a superconductor-insulator transition at a critical wave impedance \cite{bradley_quantum_1984,korshunov_effect_1989,bobbert_phase_1990,bobbert_phase_1992,glazman_new_1997,choi_quantum_1998,fazio_tunneling_1996,chow_length-scale_1998,haviland_superconducting_2000,fazio_quantum_2001,bard_superconductor-insulator_2017}, and a well-understood insulating phase \cite{vogt_one-dimensional_2015,cedergren_insulating_2017}.
Yet high-impedance arrays used in quantum computing \cite{manucharyan_fluxonium_2009,masluk_microwave_2012,nguyen_fluxonium_2019,ivan_quasicharge_2020} and metrology \cite{crescini_evidence_2022} apparently evade this transition, displaying superconducting behavior deep into the nominally insulating regime \cite{kuzmin_quantum_2019}.
The absence of critical behavior in such devices is not well understood.
Here we show that, unlike the typical quantum-critical broadening scenario, in Josephson arrays temperature dramatically shifts the critical region.
This shift leads to a regime of superconductivity at high temperature, arising from the melted zero-temperature insulator.
Our results quantitatively explain the low-temperature onset of superconductivity in nominally insulating regimes, and the transition to the strongly insulating phase.
We further present, to our knowledge, the first understanding of the onset of anomalous-metallic resistance saturation \cite{kapitulnik_anomalous_2019}.
This work demonstrates a non-trivial interplay between thermal effects and quantum criticality.
A practical consequence is that, counterintuitively, the coherence of high-impedance quantum circuits is expected to be stabilized by thermal fluctuations.}

Josephson-array superinductors are characterized by a Josephson energy $E_J$, junction charging energy $E_C$, and ground charging energy $E_g$ \cite{masluk_microwave_2012}.
A common experimental strategy for avoiding insulating behavior is to make the fugacity for quantum phase slips $y \propto e^{-4 \sqrt{2 E_J/E_C}}$ small.
However, for high-impedance arrays the fugacity is always renormalized towards infinity as temperature goes to zero \cite{giamarchi_anderson_1988,bard_superconductor-insulator_2017}, resulting in insulating behavior. 
Our key insight is that long superinductors avoid this fate by operating above the melting point of the insulating phase, where the low-temperature renormalization has yet to occur, and that this results in apparent superconducting behavior.
This effect quantitatively explains the presence of superconducting behavior, resistance saturation, and transition to strongly insulating regimes in superinductors.

\begin{figure*}
	\centering
    \includegraphics[width=\textwidth]{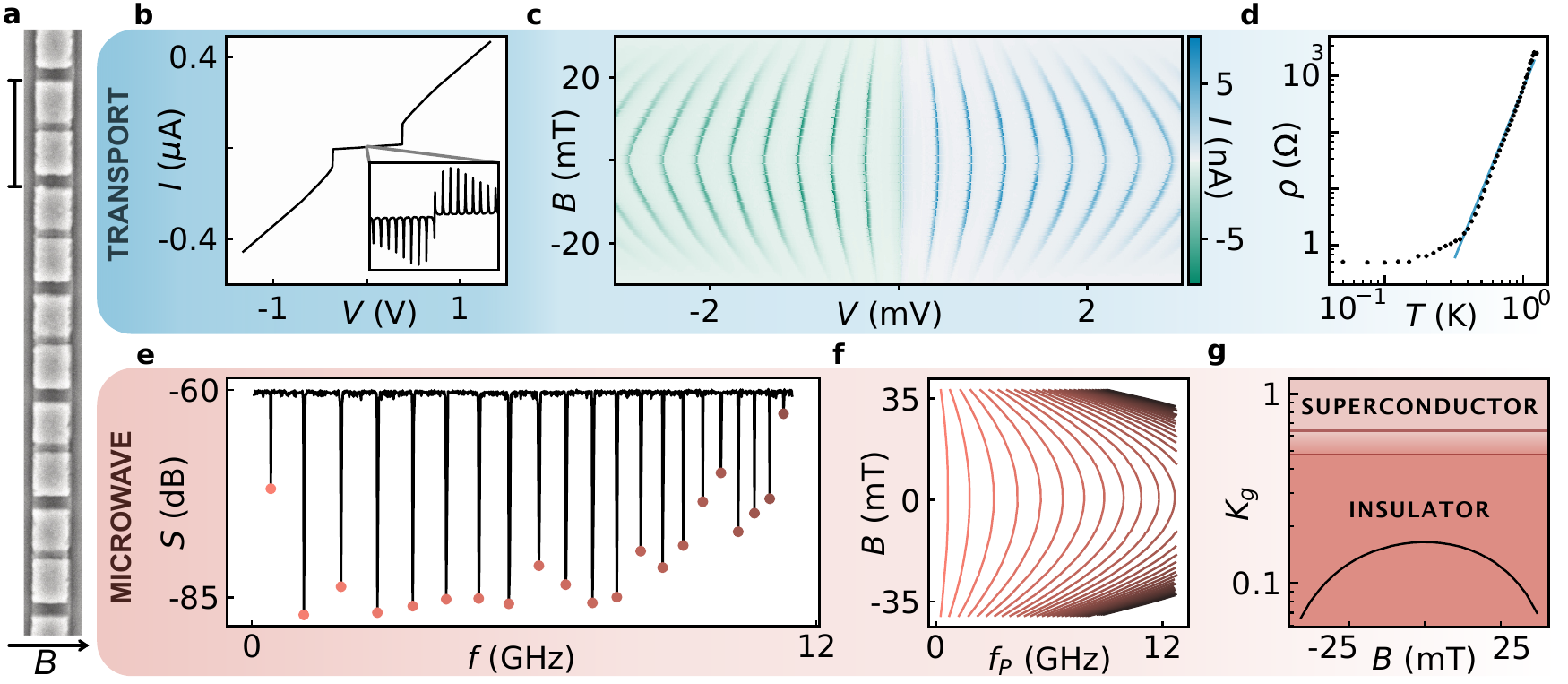}
    \caption{\textbf{Device, transport, and microwave measurement techniques.}
    \textbf{a,} Scanning electron micrograph of a small segment of the Josephson array.
    Left scale bar indicates $\mathrm{1.5\ \mu m}$.
    Arrow indicates direction of magnetic field $B$.
    \textbf{b,} Current $I$ versus source-drain bias voltage $V$.
    Inset shows small-scale current peaks over a narrow range $\mathrm{(-3,3)\ mV}$.
    \textbf{c,}~Current $I$ versus bias $V$ and magnetic field $B$ over a bias range similar to Fig.~1b inset.
    \textbf{d,} Differential resistance per junction (specific resistance) $\rho$ versus cryostat temperature $T$ measured at $V=0$ and $B=0$.
    Blue line shows power-law fit.
    $\rho$ reflects the resistance associated with the zero-bias superconducting branch, found by measuring the two-probe resistance, subtracting off four-probe-measured line resistance, and then dividing by number of junctions.
    \textbf{e,} Two-tone microwave spectroscopy.
    Probe tone transmission $S$ versus pump-tone frequency $f$,  with probe tone frequency fixed to resonance at approximately $\mathrm{6.11\ GHz}$.
    Extracted plasma-mode resonant frequencies $f_P$ indicated by colored markers.
    \textbf{f,} Evolution of measured plasma-mode frequencies $f_P$ with applied magnetic field $B$.
    \textbf{g,} Superfluid stiffness $K_g=\sqrt{E_J/(2 E_g)}$, experimentally inferred from plasma modes in \textbf{f}, versus $B$ (black line).
    Theoretically expected superconducting and insulating regimes labeled, and demarcated by a band covering the clean \cite{sondhi_continuous_1997} and dirty \cite{giamarchi_anderson_1988} limits.
    }
	\label{fig:tr_mw_intro}
\end{figure*}

Two nearly identical devices are studied: one galvanically coupled to electrical leads permitting the measurement of resistance, and one capacitively coupled to microwave transmission lines permitting the measurement of plasma modes \cite{masluk_microwave_2012,kuzmin_quantum_2019}.
Both devices consist of an array of approximately 1220 Josephson junctions fabricated using electron-beam lithography and a standard shadow evaporation process on high-resistivity silicon substrates (Fig.~\ref{fig:tr_mw_intro}a) \cite{supplement_TAS}.
For nanofabrication reasons the array islands have alternating thickness, which, in the presence of magnetic field, should give rise to an alternating gap structure while maintaining a uniform Josephson energy throughout the chain.
At zero magnetic field, each junction has nominally identical $E_J/h \approx \mathrm{76\ GHz}$, $E_g/h \approx \mathrm{1400\ GHz}$, and $E_C/h \approx \mathrm{5\ GHz}$.
These parameters are determined from analyzing microwave ($E_J,E_g$) and transport ($E_C$) measurements with several consistency checks, as described below and in the Supplement \cite{supplement_TAS}.

The working principle of the experiment is to leverage the complementary strengths of low-frequency electrical transport and microwave-domain circuit quantum electrodynamics.
These techniques differ by nine orders of magnitude in characteristic frequency, and combine to give access to both the scaling behavior, associated with low energies (transport), and the microscopic system parameters, associated with high energies (microwave).

In the transport device, a linear current-voltage characteristic at large applied voltage bias gives way to a high resistance region at low bias, whose extent is approximately given by the number of junctions $N$ times twice the superconducting gap $\Delta$ (Fig.~\ref{fig:tr_mw_intro}b).
Over a smaller range of applied voltage a series of evenly spaced current peaks are observed with an apparent supercurrent at zero bias (Fig.~\ref{fig:tr_mw_intro}b inset).
The successive current peaks can be qualitatively understood within a picture of successive voltage drops across $N$ voltage-biased Josephson junctions, with low current on the quasiparticle branches and high current when bias is a multiple of $2 \Delta/e$ \cite{haviland_superconducting_2000}.

Increasing magnetic field $B$ parallel to the chip plane suppresses supercurrent, suggesting a field-driven transition from a superconducting to an insulating state (Fig.~\ref{fig:tr_mw_intro}c).
The spacing between current peaks also decreases with $B$, indicating a reduction in the superconducting gap with magnetic field.
In the strongly superconducting regime ($B=0$), zero-bias differential resistance per junction (specific resistance) associated with the superconducting branch decreases dramatically with cryostat temperature (Fig.~\ref{fig:tr_mw_intro}d), dropping over more than three decades before saturating to a low value of $< \mathrm{1\ \Omega}$ per junction.
Due to the long length of the array, we rule out finite-size effects as a possible origin of the low-temperature saturation~\footnote{At $150~\mathrm{mK}$ the thermal length is more than an order of magnitude shorter than the device length.}.
The precipitous drop in resistance at low temperature and supercurrent features in nonlinear transport give a preliminary indication of the dominance of superconducting behavior.
We will develop a framework for understanding the behavior of specific resistance in detail, but first turn to the complementary use of microwave techniques to independently determine system parameters.

Microwave spectroscopy is performed by monitoring the transmission of a weak probe signal while the frequency of a strong pump tone is varied \cite{masluk_microwave_2012}.
A series of sharp dips are observed in probe-tone transmission $S$ (Fig.~\ref{fig:tr_mw_intro}e), corresponding to plasma modes of the array. 
The plasma modes are evenly spaced at low frequency, reflecting the speed of light and length of the array, and are clustered at high frequency due to proximity with the single-junction plasma frequency.
A simple fitting procedure allows extraction of the array parameters from the microwave data \cite{supplement_TAS}.
Performing two-tone spectroscopy as a function of field (Fig.~\ref{fig:tr_mw_intro}f), the array parameters $E_g$, $E_C$, and $E_J(B)$ are fully characterized as a function of magnetic field.
With these values fixed experimentally, it is straightforward to perform parameter-free comparisons with the theory of the superconductor-insulator transition in one dimension.

Performing this comparison (Fig.~\ref{fig:tr_mw_intro}g) reveals that the array's superfluid stiffness $K_g = \sqrt{E_J/(2 E_g)}$ is as much as an order of magnitude below the critical value for insulating behavior  \cite{sondhi_continuous_1997,giamarchi_anderson_1988}, in contrast to the observed superconducting behavior in transport.
Thus, combining the transport and microwave measurements reveals an apparent conflict with basic expectations for the superconductor-insulator phase transition.
Resolving this conflict is the central subject of this work.

\begin{figure}
	\centering
    \includegraphics[width=0.5\textwidth]{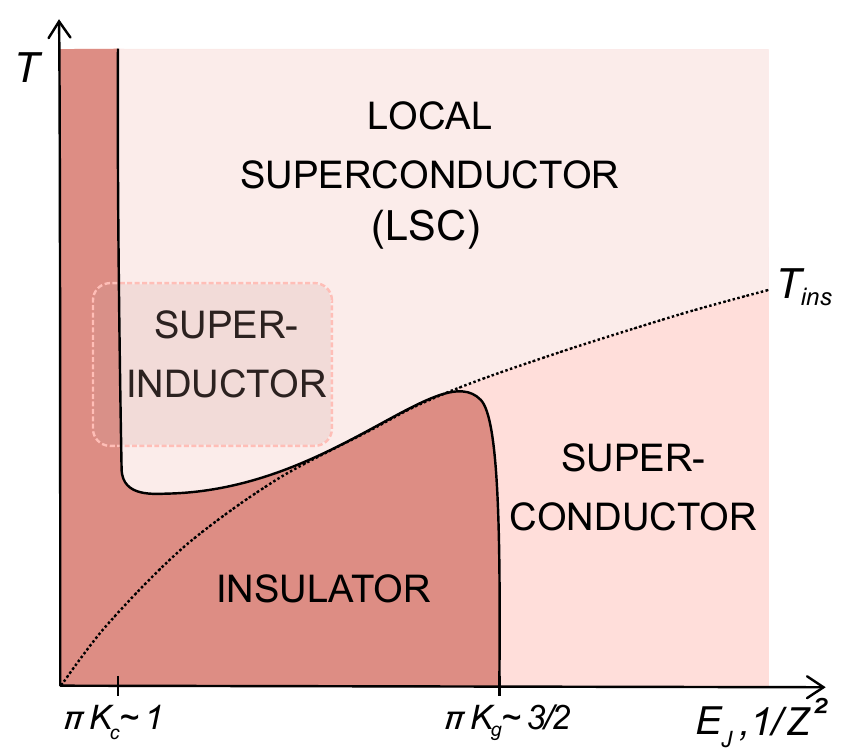}
    \caption{\textbf{Proposed phase diagram.} 
    Map of superconducting and insulating states as a function of Josephson energy $E_J$ and temperature $T$.
    Dashed line marks the boundary between long-range and short-range behavior, $T_\mathrm{ins}$, given by Eq.~\ref{eq:crossover}.
    Below $T_\mathrm{ins}$, physics is governed by the long-range superfluid stiffness $K_g$ with a superconductor-insulator transition at $\pi K_g \sim 3/2$.
    Above $T_\mathrm{ins}$, physics is governed by the short-range superfluid stiffness $K_C$ with a superconductor-insulator transition as $\pi K_C \sim 1$.
    Solid black curve traces the crossover from local to global superconductor-insulator transition.
    Outlined box indicates superinductance region probed in this experiment.
    }
	\label{fig:theory_pdiagram}
\end{figure}

The theoretical picture for understanding our observations was developed in Ref.~\cite{bard_superconductor-insulator_2017}.
Near the superconductor-insulator transition, thermal fluctuations are controlled by the timescale $\tau = h/k_B T$ and the associated thermal length $l_\mathrm{th} = v \tau$, where $v$ is a characteristic velocity with dimensions of unit cells per time.
$l_\mathrm{th}$ must be compared with the electrostatic screening length in units of unit cells, $\Lambda=\sqrt{E_g/E_C}$.
At high temperature ($l_\mathrm{th}<\Lambda$) the system is governed by the local superfluid stiffness, $K_C = \sqrt{ E_J / ( 2 E_C ) }$.
In contrast, at low temperature ($l_\mathrm{th}>\Lambda$) the system is governed by the long-range superfluid stiffness $K_g$, as assumed by standard theories of the superconductor-insulator transition.
In the superinductor limit superconductivity is locally stiff, $K_C \gg K_g$, which results in a curious regime of local superconductivity that arises from a melted $T=0$ insulator (Fig.~\ref{fig:theory_pdiagram}).
The ``melting point" of the insulator, above which local superconductivity dominates, is 
\begin{equation}
    \label{eq:crossover}
    T_\mathrm{ins} \sim \sqrt{2 E_J E_C}/\Lambda.
\end{equation} 
In the locally superconducting regime, we find that the high-temperature behavior of the specific resistance follows a power law
\begin{equation}
    \label{eq:plaw}
    \rho = \rho_0 (T/T_\mathrm{p})^{\pi K_C - 1},
\end{equation}
where $T_{\mathrm{p}} = \sqrt{2E_JE_C}/k_B$ is the plasma temperature \cite{supplement_TAS}.
Local superconductivity gives way to insulating behavior when $\pi K_C \sim 1$.
In contrast, in the low-temperature limit the power law is $2 \pi K_g - 3$, which yields the typical superconductor-insulator prediction $\pi K_g \sim 3/2$.

\begin{figure}
	\centering
    \includegraphics[width=0.5\textwidth]{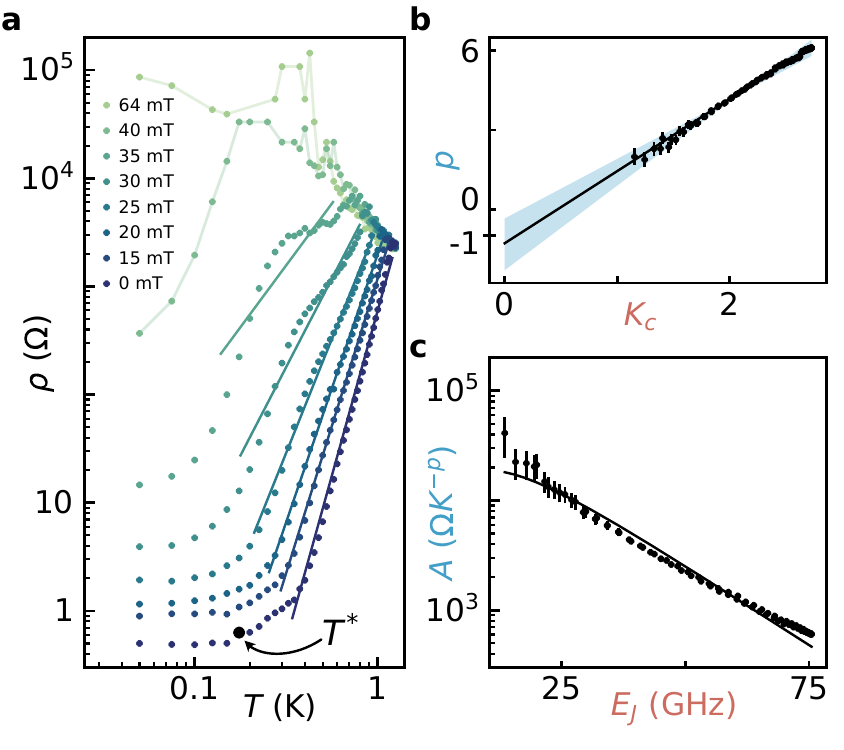}
    \caption{\textbf{Power law nature of local superconductivity.}
    \textbf{a,} Zero-bias specific differential resistance $\rho$ as a function of temperature $T$, at various magnetic fields.
    Solid lines are fits to power law expression $\rho = A T^p$.
    $T^*$ is the crossover temperature from power law to saturation behavior, extracted from the point where specific resistance goes $20\%$ above its minimum value.
    \textbf{b,} Exponent $p$ from power-law fits to transport data in \textbf{a} versus the local superfluid stiffness $K_C$ from microwave measurements.
    Solid line is a linear fit.
    Shaded blue region in \textbf{b} depicts the systematic error resulting from the choice of lower resistance cutoff in the power law fits \cite{supplement_TAS}.
    \textbf{c,} Amplitude $A$ from power-law fits to transport data versus Josephson energy $E_J$ from microwave measurements. 
    }
	\label{fig:plaw_amp}
\end{figure}

The experimentally studied devices have, at $\mathrm{B = 0}$, $\pi K_g < 1 < \pi K_C$, and $T_\mathrm{ins}\sim \mathrm{70\ mK}$, giving an initial suggestion that they are governed by local superconductivity even at low temperatures.
This hypothesis can be tested by comparing experimental measurements of temperature dependent specific resistance, $\rho(T)$, with the predicted power law in Eq.~\ref{eq:plaw}.
As shown in Fig.~\ref{fig:plaw_amp}a, increasing magnetic field weakens the temperature dependence of the specific resistance, eventually giving way to a superconductor insulator transition at high magnetic field ($B\sim \mathrm{44\ mT}$).
Fitting each specific resistance curve to a power law $\rho = A T^p$ indicates that, on the superconducting side, the exponent $p$ steadily decreases with field.
Comparing $p$ from the transport measurements with the local superfluid stiffness $K_C$ inferred from microwave measurements reveals a linear behavior (Fig.~\ref{fig:plaw_amp}b) with slope  $\mathrm{2.7 \pm 0.50}$ and intercept of $\mathrm{-1.3 \pm 1.0}$, in agreement with the predicted slope $\pi$ and intercept $-1$ for local superconductivity from Eq.~\ref{eq:plaw}, $p=\pi K_C -1$ \footnote{Parameter uncertainties are propagated from systematic bands in Fig.~\ref{fig:plaw_amp}, see Supplement \cite{supplement_TAS}.}.
We note that, near the superconductor-insulator transition, power-law behavior is interrupted by a shoulder-like feature at high temperature, which is not understood.
Amplitude dependence on $E_J$ (Fig.~\ref{fig:plaw_amp}c) is also in reasonable agreement with the prediction of Eq.~\ref{eq:plaw}, $A = \rho_0 / T_\mathrm{p}^{\pi K_C - 1}$, with a single free parameter, $\rho_0 = 4.8 \pm 0.30~\mathrm{k\Omega}$, which is of the order of single-junction normal-state resistance.
Experimental agreement with Eq.~\ref{eq:plaw} gives strong evidence that superconductivity is local, and resolves the apparent paradox of superconductivity at low $K_g$ suggested by Fig.~\ref{fig:tr_mw_intro}g.

\begin{figure}
	\centering
    \includegraphics[width=0.5\textwidth]{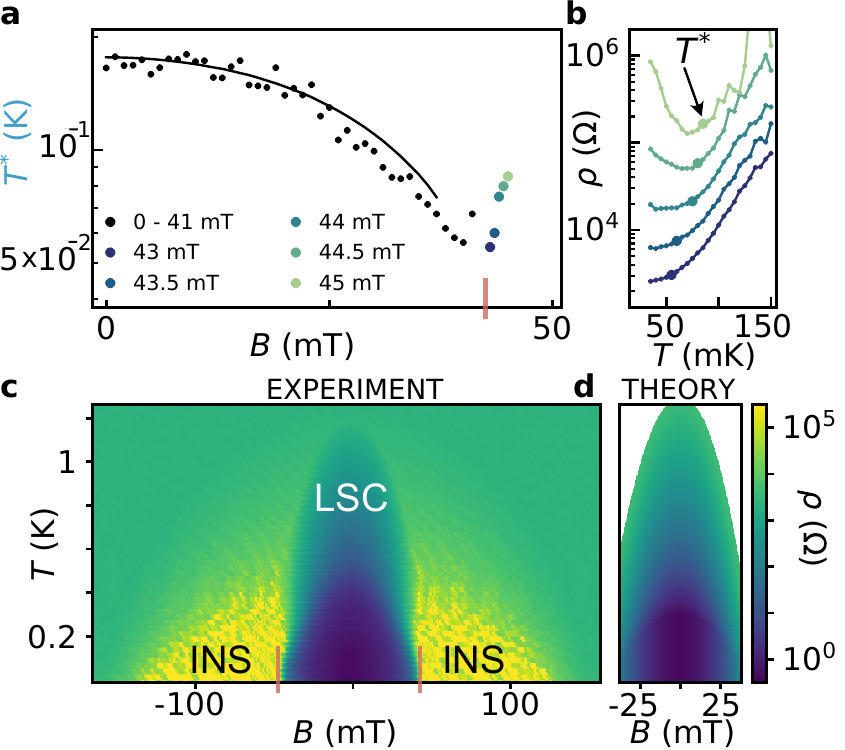}
    \caption{\textbf{Crossover physics and phase diagram.}
    \textbf{a,} Crossover temperature $T^*$ versus magnetic field $B$.
    Black line is $T^* \propto T_\mathrm{ins}$ with a proportionality constant of 2.2.
    Colored markers indicate crossover temperature at higher fields from dataset in \textbf{b}.
    Red vertical line indicates $\pi K_C = 1$, where local superconductor-insulator transition is expected.
    \textbf{b,} Zero-bias differential specific resistance $\rho$ versus temperature $T$ at higher magnetic fields, measured with higher excitation voltage and more averaging than in Fig.~\ref{fig:plaw_amp}.
    \textbf{c,} $\rho$ versus temperature $T$ and magnetic field $B$. 
    Dome of local superconductivity (LSC), and wings of insulating behavior (INS) labeled.
    Red vertical lines indicate $\pi K_C = 1$, where local superconductor-insulator transition is expected.
    \textbf{d,} Calculated specific resistance $\rho$ as a function of temperature $T$ and magnetic field $B$.
    }
	\label{fig:tcorner_pdiagram}
\end{figure}

The boundaries of local superconductivity can also be understood within the picture of Fig.~\ref{fig:theory_pdiagram}.
At low temperatures, the experimentally observed power-law behavior in resistance saturates at a crossover temperature $T^*$ (indicated in Fig.~\ref{fig:plaw_amp}a).
The crossover temperature decreases with magnetic field, as shown in Fig.~\ref{fig:tcorner_pdiagram}a, agreeing with the expected square-root dependence for $T^* \propto T_\mathrm{ins}$, which supports the view that the low-temperature saturation is in fact a crossover into the insulating state.
At high magnetic fields corresponding to $\pi K_C > 1$, $T^*$ increases with magnetic field (Fig.~\ref{fig:tcorner_pdiagram}b), consistent with a superconductor-insulator transition entering into the non-perturbative insulating regime of Ref.~\cite{bard_superconductor-insulator_2017}, where the phase-slip fugacity, $\propto e^{-8\sqrt{E_J/E_C}}$, is no longer small.
We caution that the experimental interpretation of $T^*$ is complicated for two reasons.
First, although we have performed normal-state electron thermometry and radiation thermometry and found that all characteristic temperatures are below $T^*$, thermalization at the actual superconductor-insulator transition is difficult to verify directly.
Second, different metrics for $T^*$ give quantitatively different scaling with $B$, although the decreasing trend predicted by Eq.~\ref{eq:crossover} is a robust feature.

The complete behavior of the Josephson array can be summarized by measuring a resistance ``phase diagram.''
Mapping zero-bias differential resistance as a function of magnetic field and temperature reveals a characteristic dome at low field, already identified from the power-law analysis as a local superconductor, giving way to a high-resistance insulating phase as magnetic field is increased (Fig.~\ref{fig:tcorner_pdiagram}c). 
The low-temperature boundary between superconducting and insulating states occurs at $\pi K_C \sim 1$, as expected.
We speculate that the high-field boundary of the high-resistance regime corresponds to the upper critical field of the thinnest islands of the array.

The local superconducting dome and its boundaries can be quantitatively modeled as follows.
The thermal boundary of the dome is $T=T_\mathrm{p}$, the upper cutoff scale of our renormalization-group approach \cite{bard_superconductor-insulator_2017}.
For $\alpha T_\mathrm{ins} < T < T_\mathrm{p}$ Eq.~\ref{eq:plaw} applies, with $\rho_0$ from Fig.~\ref{fig:plaw_amp}c.
For $T < \alpha T_\mathrm{ins}$ resistance saturates due to a crossover into the insulating regime, and would presumably increase at lower, experimentally inaccessible temperatures.
The constant $\alpha=5$, which tunes the crossover to insulating behavior in the model, is fixed from the experimentally observed saturation resistance at $B=0$ and in reasonable agreement with the constant found in Fig.~\ref{fig:tcorner_pdiagram}a.
For sufficiently large $B$ one approaches $\pi K_C = 1$, which sets the magnetic field boundaries of the dome.
Calculating $\rho$ according to this procedure results in a local superconducting dome in satisfactory agreement to the experiment (Fig.~\ref{fig:tcorner_pdiagram}d).
This gives evidence that the presence of local superconductivity, and its proximity to insulating phases, is well understood.

Summarizing, by combining transport and microwave measurements, we have uncovered strong evidence for a locally superconducting state in Josephson arrays arising from a $T=0$ insulator. 
This resolves the problem of apparent superconductivity in nominally insulating regimes, and clarifies where superconductor-insulator transitions are actually observed in experiment.
Our work sheds new light on the observation of high-quality microwave response in the nominally insulating regime of superinductors \cite{kuzmin_quantum_2019}, suggesting effects in addition to high-frequency mechanisms that have been previously discussed \cite{bard_decay_2018,houzet_microwave_2019,wu_theory_2019}.
Such devices operate near the ``sweet spot'' $T \approx T_\mathrm{ins}$ where temperature is low enough for well-developed local superconductivity, yet high enough to melt insulating behavior.
As a consequence, we suggest that the performance of some high-impedance quantum devices \cite{nguyen_fluxonium_2019,ivan_quasicharge_2020,winkel_transmon_2020} is actually improved by thermal fluctuations.
It is also interesting to consider if experimental studies of insulating behavior in resistively shunted Josephson junctions \cite{ryuta_phase_1997,penttila_superconductor-insulator_1999,kuzmin_coulomb_1991,murani_absence_2020} could be understood by carefully considering the role of non-zero temperature, finite-size, or non-perturbative effects \cite{masuki_absence_2022}.

Viewed from the broader perspective of response functions near quantum criticality, we have demonstrated a rare example where the thermal fluctuations with timescale $\tau=h/k_B T$ can be quantitatively traced through to experimentally measured resistance \cite{bard_superconductor-insulator_2017}.
This does not result in an effectively Planckian scattering~\cite{supplement_TAS}, as was recently observed in a different superconductor-insulator system \cite{yang_signatures_2022}.
It is also interesting to note that our saturating specific resistance curves empirically bear a strong resemblance to the anomalous-metallic phase in two-dimensional systems \cite{kapitulnik_anomalous_2019}.
In our case, saturation is understood as a crossover effect towards insulating behavior.
It would be interesting to perform a similar experimental program on a known anomalous-metallic system to test if saturation can be understood as a similar crossover effect.

\bibliography{main}

\begin{thebibliography}{45}%
  \makeatletter
  \providecommand \@ifxundefined [1]{%
   \@ifx{#1\undefined}
  }%
  \providecommand \@ifnum [1]{%
   \ifnum #1\expandafter \@firstoftwo
   \else \expandafter \@secondoftwo
   \fi
  }%
  \providecommand \@ifx [1]{%
   \ifx #1\expandafter \@firstoftwo
   \else \expandafter \@secondoftwo
   \fi
  }%
  \providecommand \natexlab [1]{#1}%
  \providecommand \enquote  [1]{``#1''}%
  \providecommand \bibnamefont  [1]{#1}%
  \providecommand \bibfnamefont [1]{#1}%
  \providecommand \citenamefont [1]{#1}%
  \providecommand \href@noop [0]{\@secondoftwo}%
  \providecommand \href [0]{\begingroup \@sanitize@url \@href}%
  \providecommand \@href[1]{\@@startlink{#1}\@@href}%
  \providecommand \@@href[1]{\endgroup#1\@@endlink}%
  \providecommand \@sanitize@url [0]{\catcode `\\12\catcode `\$12\catcode
    `\&12\catcode `\#12\catcode `\^12\catcode `\_12\catcode `\%12\relax}%
  \providecommand \@@startlink[1]{}%
  \providecommand \@@endlink[0]{}%
  \providecommand \url  [0]{\begingroup\@sanitize@url \@url }%
  \providecommand \@url [1]{\endgroup\@href {#1}{\urlprefix }}%
  \providecommand \urlprefix  [0]{URL }%
  \providecommand \Eprint [0]{\href }%
  \providecommand \doibase [0]{https://doi.org/}%
  \providecommand \selectlanguage [0]{\@gobble}%
  \providecommand \bibinfo  [0]{\@secondoftwo}%
  \providecommand \bibfield  [0]{\@secondoftwo}%
  \providecommand \translation [1]{[#1]}%
  \providecommand \BibitemOpen [0]{}%
  \providecommand \bibitemStop [0]{}%
  \providecommand \bibitemNoStop [0]{.\EOS\space}%
  \providecommand \EOS [0]{\spacefactor3000\relax}%
  \providecommand \BibitemShut  [1]{\csname bibitem#1\endcsname}%
  \let\auto@bib@innerbib\@empty
  \bibitem [{\citenamefont {Sachdev}\ and\ \citenamefont
    {Keimer}(2011)}]{sachdev_quantum_2011}%
    \BibitemOpen
    \bibfield  {author} {\bibinfo {author} {\bibfnamefont {S.}~\bibnamefont
    {Sachdev}}\ and\ \bibinfo {author} {\bibfnamefont {B.}~\bibnamefont
    {Keimer}},\ }\bibfield  {title} {\bibinfo {title} {Quantum criticality},\
    }\href {https://doi.org/10.1063/1.3554314} {\bibfield  {journal} {\bibinfo
    {journal} {Physics Today}\ }\textbf {\bibinfo {volume} {64}},\ \bibinfo
    {pages} {29} (\bibinfo {year} {2011})}\BibitemShut {NoStop}%
  \bibitem [{\citenamefont {Sondhi}\ \emph {et~al.}(1997)\citenamefont {Sondhi},
    \citenamefont {Girvin}, \citenamefont {Carini},\ and\ \citenamefont
    {Shahar}}]{sondhi_continuous_1997}%
    \BibitemOpen
    \bibfield  {author} {\bibinfo {author} {\bibfnamefont {S.~L.}\ \bibnamefont
    {Sondhi}}, \bibinfo {author} {\bibfnamefont {S.~M.}\ \bibnamefont {Girvin}},
    \bibinfo {author} {\bibfnamefont {J.~P.}\ \bibnamefont {Carini}},\ and\
    \bibinfo {author} {\bibfnamefont {D.}~\bibnamefont {Shahar}},\ }\bibfield
    {title} {\bibinfo {title} {Continuous quantum phase transitions},\ }\href
    {https://doi.org/10.1103/RevModPhys.69.315} {\bibfield  {journal} {\bibinfo
    {journal} {Rev. Mod. Phys.}\ }\textbf {\bibinfo {volume} {69}},\ \bibinfo
    {pages} {315} (\bibinfo {year} {1997})}\BibitemShut {NoStop}%
  \bibitem [{\citenamefont {Bradley}\ and\ \citenamefont
    {Doniach}(1984)}]{bradley_quantum_1984}%
    \BibitemOpen
    \bibfield  {author} {\bibinfo {author} {\bibfnamefont {R.~M.}\ \bibnamefont
    {Bradley}}\ and\ \bibinfo {author} {\bibfnamefont {S.}~\bibnamefont
    {Doniach}},\ }\bibfield  {title} {\bibinfo {title} {Quantum fluctuations in
    chains of josephson junctions},\ }\href
    {https://doi.org/10.1103/PhysRevB.30.1138} {\bibfield  {journal} {\bibinfo
    {journal} {Phys. Rev. B}\ }\textbf {\bibinfo {volume} {30}},\ \bibinfo
    {pages} {1138} (\bibinfo {year} {1984})}\BibitemShut {NoStop}%
  \bibitem [{\citenamefont {Korshunov}(1989)}]{korshunov_effect_1989}%
    \BibitemOpen
    \bibfield  {author} {\bibinfo {author} {\bibfnamefont {S.~E.}\ \bibnamefont
    {Korshunov}},\ }\bibfield  {title} {\bibinfo {title} {Effect of dissipation
    on the low-temperature properties of a tunnel-junction chain},\ }\href@noop
    {} {\bibfield  {journal} {\bibinfo  {journal} {Sov. Phys. JETP}\ }\textbf
    {\bibinfo {volume} {68}},\ \bibinfo {pages} {609} (\bibinfo {year}
    {1989})}\BibitemShut {NoStop}%
  \bibitem [{\citenamefont {Bobbert}\ \emph {et~al.}(1990)\citenamefont
    {Bobbert}, \citenamefont {Fazio}, \citenamefont {Sch\"on},\ and\
    \citenamefont {Zimanyi}}]{bobbert_phase_1990}%
    \BibitemOpen
    \bibfield  {author} {\bibinfo {author} {\bibfnamefont {P.~A.}\ \bibnamefont
    {Bobbert}}, \bibinfo {author} {\bibfnamefont {R.}~\bibnamefont {Fazio}},
    \bibinfo {author} {\bibfnamefont {G.}~\bibnamefont {Sch\"on}},\ and\ \bibinfo
    {author} {\bibfnamefont {G.~T.}\ \bibnamefont {Zimanyi}},\ }\bibfield
    {title} {\bibinfo {title} {Phase transitions in dissipative josephson
    chains},\ }\href {https://doi.org/10.1103/PhysRevB.41.4009} {\bibfield
    {journal} {\bibinfo  {journal} {Phys. Rev. B}\ }\textbf {\bibinfo {volume}
    {41}},\ \bibinfo {pages} {4009} (\bibinfo {year} {1990})}\BibitemShut
    {NoStop}%
  \bibitem [{\citenamefont {Bobbert}\ \emph {et~al.}(1992)\citenamefont
    {Bobbert}, \citenamefont {Fazio}, \citenamefont {Sch\"on},\ and\
    \citenamefont {Zaikin}}]{bobbert_phase_1992}%
    \BibitemOpen
    \bibfield  {author} {\bibinfo {author} {\bibfnamefont {P.~A.}\ \bibnamefont
    {Bobbert}}, \bibinfo {author} {\bibfnamefont {R.}~\bibnamefont {Fazio}},
    \bibinfo {author} {\bibfnamefont {G.}~\bibnamefont {Sch\"on}},\ and\ \bibinfo
    {author} {\bibfnamefont {A.~D.}\ \bibnamefont {Zaikin}},\ }\bibfield  {title}
    {\bibinfo {title} {Phase transitions in dissipative josephson chains: Monte
    carlo results and response functions},\ }\href
    {https://doi.org/10.1103/PhysRevB.45.2294} {\bibfield  {journal} {\bibinfo
    {journal} {Phys. Rev. B}\ }\textbf {\bibinfo {volume} {45}},\ \bibinfo
    {pages} {2294} (\bibinfo {year} {1992})}\BibitemShut {NoStop}%
  \bibitem [{\citenamefont {Glazman}\ and\ \citenamefont
    {Larkin}(1997)}]{glazman_new_1997}%
    \BibitemOpen
    \bibfield  {author} {\bibinfo {author} {\bibfnamefont {L.~I.}\ \bibnamefont
    {Glazman}}\ and\ \bibinfo {author} {\bibfnamefont {A.~I.}\ \bibnamefont
    {Larkin}},\ }\bibfield  {title} {\bibinfo {title} {New quantum phase in a
    one-dimensional josephson array},\ }\href
    {https://doi.org/10.1103/PhysRevLett.79.3736} {\bibfield  {journal} {\bibinfo
     {journal} {Phys. Rev. Lett.}\ }\textbf {\bibinfo {volume} {79}},\ \bibinfo
    {pages} {3736} (\bibinfo {year} {1997})}\BibitemShut {NoStop}%
  \bibitem [{\citenamefont {Choi}\ \emph {et~al.}(1998)\citenamefont {Choi},
    \citenamefont {Yi}, \citenamefont {Choi}, \citenamefont {Choi},\ and\
    \citenamefont {Lee}}]{choi_quantum_1998}%
    \BibitemOpen
    \bibfield  {author} {\bibinfo {author} {\bibfnamefont {M.-S.}\ \bibnamefont
    {Choi}}, \bibinfo {author} {\bibfnamefont {J.}~\bibnamefont {Yi}}, \bibinfo
    {author} {\bibfnamefont {M.~Y.}\ \bibnamefont {Choi}}, \bibinfo {author}
    {\bibfnamefont {J.}~\bibnamefont {Choi}},\ and\ \bibinfo {author}
    {\bibfnamefont {S.-I.}\ \bibnamefont {Lee}},\ }\bibfield  {title} {\bibinfo
    {title} {Quantum phase transitions in josephson-junction chains},\ }\href
    {https://doi.org/10.1103/PhysRevB.57.R716} {\bibfield  {journal} {\bibinfo
    {journal} {Phys. Rev. B}\ }\textbf {\bibinfo {volume} {57}},\ \bibinfo
    {pages} {R716} (\bibinfo {year} {1998})}\BibitemShut {NoStop}%
  \bibitem [{\citenamefont {Fazio}\ \emph {et~al.}(1996)\citenamefont {Fazio},
    \citenamefont {Wagenblast}, \citenamefont {Winkelholz},\ and\ \citenamefont
    {Schön}}]{fazio_tunneling_1996}%
    \BibitemOpen
    \bibfield  {author} {\bibinfo {author} {\bibfnamefont {R.}~\bibnamefont
    {Fazio}}, \bibinfo {author} {\bibfnamefont {K.-H.}\ \bibnamefont
    {Wagenblast}}, \bibinfo {author} {\bibfnamefont {C.}~\bibnamefont
    {Winkelholz}},\ and\ \bibinfo {author} {\bibfnamefont {G.}~\bibnamefont
    {Schön}},\ }\bibfield  {title} {\bibinfo {title} {Tunneling into
    one-dimensional josephson chains and luttinger liquids},\ }\href
    {https://doi.org/https://doi.org/10.1016/0921-4526(96)00219-0} {\bibfield
    {journal} {\bibinfo  {journal} {Physica B: Condensed Matter}\ }\textbf
    {\bibinfo {volume} {222}},\ \bibinfo {pages} {364} (\bibinfo {year}
    {1996})},\ \bibinfo {note} {proceedings of the ICTP Workshop on Josephson
    Junction Arrays}\BibitemShut {NoStop}%
  \bibitem [{\citenamefont {Chow}\ \emph {et~al.}(1998)\citenamefont {Chow},
    \citenamefont {Delsing},\ and\ \citenamefont
    {Haviland}}]{chow_length-scale_1998}%
    \BibitemOpen
    \bibfield  {author} {\bibinfo {author} {\bibfnamefont {E.}~\bibnamefont
    {Chow}}, \bibinfo {author} {\bibfnamefont {P.}~\bibnamefont {Delsing}},\ and\
    \bibinfo {author} {\bibfnamefont {D.~B.}\ \bibnamefont {Haviland}},\
    }\bibfield  {title} {\bibinfo {title} {Length-scale dependence of the
    superconductor-to-insulator quantum phase transition in one dimension},\
    }\href {https://doi.org/10.1103/PhysRevLett.81.204} {\bibfield  {journal}
    {\bibinfo  {journal} {Phys. Rev. Lett.}\ }\textbf {\bibinfo {volume} {81}},\
    \bibinfo {pages} {204} (\bibinfo {year} {1998})}\BibitemShut {NoStop}%
  \bibitem [{\citenamefont {Haviland}\ \emph {et~al.}(2000)\citenamefont
    {Haviland}, \citenamefont {Andersson},\ and\ \citenamefont
    {Ågren}}]{haviland_superconducting_2000}%
    \BibitemOpen
    \bibfield  {author} {\bibinfo {author} {\bibfnamefont {D.~B.}\ \bibnamefont
    {Haviland}}, \bibinfo {author} {\bibfnamefont {K.}~\bibnamefont
    {Andersson}},\ and\ \bibinfo {author} {\bibfnamefont {P.}~\bibnamefont
    {Ågren}},\ }\bibfield  {title} {\bibinfo {title} {Superconducting and
    insulating behavior in one-dimensional josephson junction arrays},\ }\href
    {https://doi.org/10.1023/A:1004603814529} {\bibfield  {journal} {\bibinfo
    {journal} {Journal of Low Temperature Physics}\ }\textbf {\bibinfo {volume}
    {118}},\ \bibinfo {pages} {733} (\bibinfo {year} {2000})}\BibitemShut
    {NoStop}%
  \bibitem [{\citenamefont {Fazio}\ and\ \citenamefont {{van der
    Zant}}(2001)}]{fazio_quantum_2001}%
    \BibitemOpen
    \bibfield  {author} {\bibinfo {author} {\bibfnamefont {R.}~\bibnamefont
    {Fazio}}\ and\ \bibinfo {author} {\bibfnamefont {H.}~\bibnamefont {{van der
    Zant}}},\ }\bibfield  {title} {\bibinfo {title} {Quantum phase transitions
    and vortex dynamics in superconducting networks},\ }\href
    {https://doi.org/https://doi.org/10.1016/S0370-1573(01)00022-9} {\bibfield
    {journal} {\bibinfo  {journal} {Physics Reports}\ }\textbf {\bibinfo {volume}
    {355}},\ \bibinfo {pages} {235} (\bibinfo {year} {2001})}\BibitemShut
    {NoStop}%
  \bibitem [{\citenamefont {Bard}\ \emph {et~al.}(2017)\citenamefont {Bard},
    \citenamefont {Protopopov}, \citenamefont {Gornyi}, \citenamefont
    {Shnirman},\ and\ \citenamefont
    {Mirlin}}]{bard_superconductor-insulator_2017}%
    \BibitemOpen
    \bibfield  {author} {\bibinfo {author} {\bibfnamefont {M.}~\bibnamefont
    {Bard}}, \bibinfo {author} {\bibfnamefont {I.~V.}\ \bibnamefont
    {Protopopov}}, \bibinfo {author} {\bibfnamefont {I.~V.}\ \bibnamefont
    {Gornyi}}, \bibinfo {author} {\bibfnamefont {A.}~\bibnamefont {Shnirman}},\
    and\ \bibinfo {author} {\bibfnamefont {A.~D.}\ \bibnamefont {Mirlin}},\
    }\bibfield  {title} {\bibinfo {title} {Superconductor-insulator transition in
    disordered josephson-junction chains},\ }\href
    {https://doi.org/10.1103/PhysRevB.96.064514} {\bibfield  {journal} {\bibinfo
    {journal} {Phys. Rev. B}\ }\textbf {\bibinfo {volume} {96}},\ \bibinfo
    {pages} {064514} (\bibinfo {year} {2017})}\BibitemShut {NoStop}%
  \bibitem [{\citenamefont {Vogt}\ \emph {et~al.}(2015)\citenamefont {Vogt},
    \citenamefont {Sch\"afer}, \citenamefont {Rotzinger}, \citenamefont {Cui},
    \citenamefont {Fiebig}, \citenamefont {Shnirman},\ and\ \citenamefont
    {Ustinov}}]{vogt_one-dimensional_2015}%
    \BibitemOpen
    \bibfield  {author} {\bibinfo {author} {\bibfnamefont {N.}~\bibnamefont
    {Vogt}}, \bibinfo {author} {\bibfnamefont {R.}~\bibnamefont {Sch\"afer}},
    \bibinfo {author} {\bibfnamefont {H.}~\bibnamefont {Rotzinger}}, \bibinfo
    {author} {\bibfnamefont {W.}~\bibnamefont {Cui}}, \bibinfo {author}
    {\bibfnamefont {A.}~\bibnamefont {Fiebig}}, \bibinfo {author} {\bibfnamefont
    {A.}~\bibnamefont {Shnirman}},\ and\ \bibinfo {author} {\bibfnamefont
    {A.~V.}\ \bibnamefont {Ustinov}},\ }\bibfield  {title} {\bibinfo {title}
    {One-dimensional josephson junction arrays: Lifting the coulomb blockade by
    depinning},\ }\href {https://doi.org/10.1103/PhysRevB.92.045435} {\bibfield
    {journal} {\bibinfo  {journal} {Phys. Rev. B}\ }\textbf {\bibinfo {volume}
    {92}},\ \bibinfo {pages} {045435} (\bibinfo {year} {2015})}\BibitemShut
    {NoStop}%
  \bibitem [{\citenamefont {Cedergren}\ \emph {et~al.}(2017)\citenamefont
    {Cedergren}, \citenamefont {Ackroyd}, \citenamefont {Kafanov}, \citenamefont
    {Vogt}, \citenamefont {Shnirman},\ and\ \citenamefont
    {Duty}}]{cedergren_insulating_2017}%
    \BibitemOpen
    \bibfield  {author} {\bibinfo {author} {\bibfnamefont {K.}~\bibnamefont
    {Cedergren}}, \bibinfo {author} {\bibfnamefont {R.}~\bibnamefont {Ackroyd}},
    \bibinfo {author} {\bibfnamefont {S.}~\bibnamefont {Kafanov}}, \bibinfo
    {author} {\bibfnamefont {N.}~\bibnamefont {Vogt}}, \bibinfo {author}
    {\bibfnamefont {A.}~\bibnamefont {Shnirman}},\ and\ \bibinfo {author}
    {\bibfnamefont {T.}~\bibnamefont {Duty}},\ }\bibfield  {title} {\bibinfo
    {title} {Insulating josephson junction chains as pinned luttinger liquids},\
    }\href {https://doi.org/10.1103/PhysRevLett.119.167701} {\bibfield  {journal}
    {\bibinfo  {journal} {Phys. Rev. Lett.}\ }\textbf {\bibinfo {volume} {119}},\
    \bibinfo {pages} {167701} (\bibinfo {year} {2017})}\BibitemShut {NoStop}%
  \bibitem [{\citenamefont {Manucharyan}\ \emph {et~al.}(2009)\citenamefont
    {Manucharyan}, \citenamefont {Koch}, \citenamefont {Glazman},\ and\
    \citenamefont {Devoret}}]{manucharyan_fluxonium_2009}%
    \BibitemOpen
    \bibfield  {author} {\bibinfo {author} {\bibfnamefont {V.~E.}\ \bibnamefont
    {Manucharyan}}, \bibinfo {author} {\bibfnamefont {J.}~\bibnamefont {Koch}},
    \bibinfo {author} {\bibfnamefont {L.~I.}\ \bibnamefont {Glazman}},\ and\
    \bibinfo {author} {\bibfnamefont {M.~H.}\ \bibnamefont {Devoret}},\
    }\bibfield  {title} {\bibinfo {title} {Fluxonium: Single cooper-pair circuit
    free of charge offsets},\ }\href {https://doi.org/10.1126/science.1175552}
    {\bibfield  {journal} {\bibinfo  {journal} {Science}\ }\textbf {\bibinfo
    {volume} {326}},\ \bibinfo {pages} {113} (\bibinfo {year}
    {2009})}\BibitemShut {NoStop}%
  \bibitem [{\citenamefont {Masluk}\ \emph {et~al.}(2012)\citenamefont {Masluk},
    \citenamefont {Pop}, \citenamefont {Kamal}, \citenamefont {Minev},\ and\
    \citenamefont {Devoret}}]{masluk_microwave_2012}%
    \BibitemOpen
    \bibfield  {author} {\bibinfo {author} {\bibfnamefont {N.~A.}\ \bibnamefont
    {Masluk}}, \bibinfo {author} {\bibfnamefont {I.~M.}\ \bibnamefont {Pop}},
    \bibinfo {author} {\bibfnamefont {A.}~\bibnamefont {Kamal}}, \bibinfo
    {author} {\bibfnamefont {Z.~K.}\ \bibnamefont {Minev}},\ and\ \bibinfo
    {author} {\bibfnamefont {M.~H.}\ \bibnamefont {Devoret}},\ }\bibfield
    {title} {\bibinfo {title} {Microwave characterization of josephson junction
    arrays: Implementing a low loss superinductance},\ }\href
    {https://doi.org/10.1103/PhysRevLett.109.137002} {\bibfield  {journal}
    {\bibinfo  {journal} {Phys. Rev. Lett.}\ }\textbf {\bibinfo {volume} {109}},\
    \bibinfo {pages} {137002} (\bibinfo {year} {2012})}\BibitemShut {NoStop}%
  \bibitem [{\citenamefont {Nguyen}\ \emph {et~al.}(2019)\citenamefont {Nguyen},
    \citenamefont {Lin}, \citenamefont {Somoroff}, \citenamefont {Mencia},
    \citenamefont {Grabon},\ and\ \citenamefont
    {Manucharyan}}]{nguyen_fluxonium_2019}%
    \BibitemOpen
    \bibfield  {author} {\bibinfo {author} {\bibfnamefont {L.~B.}\ \bibnamefont
    {Nguyen}}, \bibinfo {author} {\bibfnamefont {Y.-H.}\ \bibnamefont {Lin}},
    \bibinfo {author} {\bibfnamefont {A.}~\bibnamefont {Somoroff}}, \bibinfo
    {author} {\bibfnamefont {R.}~\bibnamefont {Mencia}}, \bibinfo {author}
    {\bibfnamefont {N.}~\bibnamefont {Grabon}},\ and\ \bibinfo {author}
    {\bibfnamefont {V.~E.}\ \bibnamefont {Manucharyan}},\ }\bibfield  {title}
    {\bibinfo {title} {High-coherence fluxonium qubit},\ }\href
    {https://doi.org/10.1103/PhysRevX.9.041041} {\bibfield  {journal} {\bibinfo
    {journal} {Phys. Rev. X}\ }\textbf {\bibinfo {volume} {9}},\ \bibinfo {pages}
    {041041} (\bibinfo {year} {2019})}\BibitemShut {NoStop}%
  \bibitem [{\citenamefont {Pechenezhskiy}\ \emph {et~al.}(2020)\citenamefont
    {Pechenezhskiy}, \citenamefont {Mencia}, \citenamefont {Nguyen},
    \citenamefont {Lin},\ and\ \citenamefont
    {Manucharyan}}]{ivan_quasicharge_2020}%
    \BibitemOpen
    \bibfield  {author} {\bibinfo {author} {\bibfnamefont {I.~V.}\ \bibnamefont
    {Pechenezhskiy}}, \bibinfo {author} {\bibfnamefont {R.~A.}\ \bibnamefont
    {Mencia}}, \bibinfo {author} {\bibfnamefont {L.~B.}\ \bibnamefont {Nguyen}},
    \bibinfo {author} {\bibfnamefont {Y.-H.}\ \bibnamefont {Lin}},\ and\ \bibinfo
    {author} {\bibfnamefont {V.~E.}\ \bibnamefont {Manucharyan}},\ }\bibfield
    {title} {\bibinfo {title} {The superconducting quasicharge qubit},\ }\href
    {https://doi.org/10.1038/s41586-020-2687-9} {\bibfield  {journal} {\bibinfo
    {journal} {Nature}\ }\textbf {\bibinfo {volume} {585}},\ \bibinfo {pages}
    {368} (\bibinfo {year} {2020})}\BibitemShut {NoStop}%
  \bibitem [{\citenamefont {Crescini}\ \emph {et~al.}(2022)\citenamefont
    {Crescini}, \citenamefont {Cailleaux}, \citenamefont {Guichard},
    \citenamefont {Naud}, \citenamefont {Buisson}, \citenamefont {Murch},\ and\
    \citenamefont {Roch}}]{crescini_evidence_2022}%
    \BibitemOpen
    \bibfield  {author} {\bibinfo {author} {\bibfnamefont {N.}~\bibnamefont
    {Crescini}}, \bibinfo {author} {\bibfnamefont {S.}~\bibnamefont {Cailleaux}},
    \bibinfo {author} {\bibfnamefont {W.}~\bibnamefont {Guichard}}, \bibinfo
    {author} {\bibfnamefont {C.}~\bibnamefont {Naud}}, \bibinfo {author}
    {\bibfnamefont {O.}~\bibnamefont {Buisson}}, \bibinfo {author} {\bibfnamefont
    {K.}~\bibnamefont {Murch}},\ and\ \bibinfo {author} {\bibfnamefont
    {N.}~\bibnamefont {Roch}},\ }\bibfield  {title} {\bibinfo {title} {Evidence
    of dual shapiro steps in a josephson junctions array},\ }\bibfield  {journal}
    {\bibinfo  {journal} {arXiv:2207.09381}\ }\href
    {https://doi.org/10.48550/arXiv.2207.09381} {10.48550/arXiv.2207.09381}
    (\bibinfo {year} {2022})\BibitemShut {NoStop}%
  \bibitem [{\citenamefont {Kuzmin}\ \emph {et~al.}(2019)\citenamefont {Kuzmin},
    \citenamefont {Mencia}, \citenamefont {Grabon}, \citenamefont {Mehta},
    \citenamefont {Lin},\ and\ \citenamefont
    {Manucharyan}}]{kuzmin_quantum_2019}%
    \BibitemOpen
    \bibfield  {author} {\bibinfo {author} {\bibfnamefont {R.}~\bibnamefont
    {Kuzmin}}, \bibinfo {author} {\bibfnamefont {R.}~\bibnamefont {Mencia}},
    \bibinfo {author} {\bibfnamefont {N.}~\bibnamefont {Grabon}}, \bibinfo
    {author} {\bibfnamefont {N.}~\bibnamefont {Mehta}}, \bibinfo {author}
    {\bibfnamefont {Y.~H.}\ \bibnamefont {Lin}},\ and\ \bibinfo {author}
    {\bibfnamefont {V.~E.}\ \bibnamefont {Manucharyan}},\ }\bibfield  {title}
    {\bibinfo {title} {Quantum electrodynamics of a superconductor--insulator
    phase transition},\ }\href {https://doi.org/10.1038/s41567-019-0553-1}
    {\bibfield  {journal} {\bibinfo  {journal} {Nature Physics}\ }\textbf
    {\bibinfo {volume} {15}},\ \bibinfo {pages} {930} (\bibinfo {year}
    {2019})}\BibitemShut {NoStop}%
  \bibitem [{\citenamefont {Kapitulnik}\ \emph {et~al.}(2019)\citenamefont
    {Kapitulnik}, \citenamefont {Kivelson},\ and\ \citenamefont
    {Spivak}}]{kapitulnik_anomalous_2019}%
    \BibitemOpen
    \bibfield  {author} {\bibinfo {author} {\bibfnamefont {A.}~\bibnamefont
    {Kapitulnik}}, \bibinfo {author} {\bibfnamefont {S.~A.}\ \bibnamefont
    {Kivelson}},\ and\ \bibinfo {author} {\bibfnamefont {B.}~\bibnamefont
    {Spivak}},\ }\bibfield  {title} {\bibinfo {title} {Colloquium: Anomalous
    metals: Failed superconductors},\ }\href
    {https://doi.org/10.1103/RevModPhys.91.011002} {\bibfield  {journal}
    {\bibinfo  {journal} {Rev. Mod. Phys.}\ }\textbf {\bibinfo {volume} {91}},\
    \bibinfo {pages} {011002} (\bibinfo {year} {2019})}\BibitemShut {NoStop}%
  \bibitem [{\citenamefont {Giamarchi}\ and\ \citenamefont
    {Schulz}(1988)}]{giamarchi_anderson_1988}%
    \BibitemOpen
    \bibfield  {author} {\bibinfo {author} {\bibfnamefont {T.}~\bibnamefont
    {Giamarchi}}\ and\ \bibinfo {author} {\bibfnamefont {H.~J.}\ \bibnamefont
    {Schulz}},\ }\bibfield  {title} {\bibinfo {title} {Anderson localization and
    interactions in one-dimensional metals},\ }\href
    {https://doi.org/10.1103/PhysRevB.37.325} {\bibfield  {journal} {\bibinfo
    {journal} {Phys. Rev. B}\ }\textbf {\bibinfo {volume} {37}},\ \bibinfo
    {pages} {325} (\bibinfo {year} {1988})}\BibitemShut {NoStop}%
  \bibitem [{sup()}]{supplement_TAS}%
    \BibitemOpen
    \href@noop {} {\bibinfo  {journal} {Supplementary materials}\ }\BibitemShut
    {NoStop}%
  \bibitem [{Note1()}]{Note1}%
    \BibitemOpen
  \bibfield  {journal} {  }\bibinfo {note} {At $150~\protect \mathrm {mK}$ the
    thermal length is more than an order of magnitude shorter than the device
    length.}\BibitemShut {Stop}%
  \bibitem [{Note2()}]{Note2}%
    \BibitemOpen
    \bibinfo {note} {Parameter uncertainties are propagated from systematic bands
    in Fig.~\ref {fig:plaw_amp}, see Supplement \cite
    {supplement_TAS}.}\BibitemShut {Stop}%
  \bibitem [{\citenamefont {Bard}\ \emph {et~al.}(2018)\citenamefont {Bard},
    \citenamefont {Protopopov},\ and\ \citenamefont {Mirlin}}]{bard_decay_2018}%
    \BibitemOpen
    \bibfield  {author} {\bibinfo {author} {\bibfnamefont {M.}~\bibnamefont
    {Bard}}, \bibinfo {author} {\bibfnamefont {I.~V.}\ \bibnamefont
    {Protopopov}},\ and\ \bibinfo {author} {\bibfnamefont {A.~D.}\ \bibnamefont
    {Mirlin}},\ }\bibfield  {title} {\bibinfo {title} {Decay of plasmonic waves
    in josephson junction chains},\ }\href
    {https://doi.org/10.1103/PhysRevB.98.224513} {\bibfield  {journal} {\bibinfo
    {journal} {Phys. Rev. B}\ }\textbf {\bibinfo {volume} {98}},\ \bibinfo
    {pages} {224513} (\bibinfo {year} {2018})}\BibitemShut {NoStop}%
  \bibitem [{\citenamefont {Houzet}\ and\ \citenamefont
    {Glazman}(2019)}]{houzet_microwave_2019}%
    \BibitemOpen
    \bibfield  {author} {\bibinfo {author} {\bibfnamefont {M.}~\bibnamefont
    {Houzet}}\ and\ \bibinfo {author} {\bibfnamefont {L.~I.}\ \bibnamefont
    {Glazman}},\ }\bibfield  {title} {\bibinfo {title} {Microwave spectroscopy of
    a weakly pinned charge density wave in a superinductor},\ }\href
    {https://doi.org/10.1103/PhysRevLett.122.237701} {\bibfield  {journal}
    {\bibinfo  {journal} {Phys. Rev. Lett.}\ }\textbf {\bibinfo {volume} {122}},\
    \bibinfo {pages} {237701} (\bibinfo {year} {2019})}\BibitemShut {NoStop}%
  \bibitem [{\citenamefont {Wu}\ and\ \citenamefont
    {Sau}(2019)}]{wu_theory_2019}%
    \BibitemOpen
    \bibfield  {author} {\bibinfo {author} {\bibfnamefont {H.-K.}\ \bibnamefont
    {Wu}}\ and\ \bibinfo {author} {\bibfnamefont {J.~D.}\ \bibnamefont {Sau}},\
    }\bibfield  {title} {\bibinfo {title} {Theory of coherent phase modes in
    insulating josephson junction chains},\ }\href
    {https://doi.org/10.1103/PhysRevB.99.214509} {\bibfield  {journal} {\bibinfo
    {journal} {Phys. Rev. B}\ }\textbf {\bibinfo {volume} {99}},\ \bibinfo
    {pages} {214509} (\bibinfo {year} {2019})}\BibitemShut {NoStop}%
  \bibitem [{\citenamefont {Winkel}\ \emph {et~al.}(2020)\citenamefont {Winkel},
    \citenamefont {Borisov}, \citenamefont {Gr\"unhaupt}, \citenamefont {Rieger},
    \citenamefont {Spiecker}, \citenamefont {Valenti}, \citenamefont {Ustinov},
    \citenamefont {Wernsdorfer},\ and\ \citenamefont
    {Pop}}]{winkel_transmon_2020}%
    \BibitemOpen
    \bibfield  {author} {\bibinfo {author} {\bibfnamefont {P.}~\bibnamefont
    {Winkel}}, \bibinfo {author} {\bibfnamefont {K.}~\bibnamefont {Borisov}},
    \bibinfo {author} {\bibfnamefont {L.}~\bibnamefont {Gr\"unhaupt}}, \bibinfo
    {author} {\bibfnamefont {D.}~\bibnamefont {Rieger}}, \bibinfo {author}
    {\bibfnamefont {M.}~\bibnamefont {Spiecker}}, \bibinfo {author}
    {\bibfnamefont {F.}~\bibnamefont {Valenti}}, \bibinfo {author} {\bibfnamefont
    {A.~V.}\ \bibnamefont {Ustinov}}, \bibinfo {author} {\bibfnamefont
    {W.}~\bibnamefont {Wernsdorfer}},\ and\ \bibinfo {author} {\bibfnamefont
    {I.~M.}\ \bibnamefont {Pop}},\ }\bibfield  {title} {\bibinfo {title}
    {Implementation of a transmon qubit using superconducting granular
    aluminum},\ }\href {https://doi.org/10.1103/PhysRevX.10.031032} {\bibfield
    {journal} {\bibinfo  {journal} {Phys. Rev. X}\ }\textbf {\bibinfo {volume}
    {10}},\ \bibinfo {pages} {031032} (\bibinfo {year} {2020})}\BibitemShut
    {NoStop}%
  \bibitem [{\citenamefont {Yagi}\ \emph {et~al.}(1997)\citenamefont {Yagi},
    \citenamefont {Kobayashi},\ and\ \citenamefont {Ootuka}}]{ryuta_phase_1997}%
    \BibitemOpen
    \bibfield  {author} {\bibinfo {author} {\bibfnamefont {R.}~\bibnamefont
    {Yagi}}, \bibinfo {author} {\bibfnamefont {S.-i.}\ \bibnamefont
    {Kobayashi}},\ and\ \bibinfo {author} {\bibfnamefont {Y.}~\bibnamefont
    {Ootuka}},\ }\bibfield  {title} {\bibinfo {title} {Phase diagram for
    superconductor-insulator transition in single small josephson junctions with
    shunt resistor},\ }\href {https://doi.org/10.1143/JPSJ.66.3722} {\bibfield
    {journal} {\bibinfo  {journal} {Journal of the Physical Society of Japan}\
    }\textbf {\bibinfo {volume} {66}},\ \bibinfo {pages} {3722} (\bibinfo {year}
    {1997})}\BibitemShut {NoStop}%
  \bibitem [{\citenamefont {Penttil\"a}\ \emph {et~al.}(1999)\citenamefont
    {Penttil\"a}, \citenamefont {Parts}, \citenamefont {Hakonen}, \citenamefont
    {Paalanen},\ and\ \citenamefont
    {Sonin}}]{penttila_superconductor-insulator_1999}%
    \BibitemOpen
    \bibfield  {author} {\bibinfo {author} {\bibfnamefont {J.~S.}\ \bibnamefont
    {Penttil\"a}}, \bibinfo {author} {\bibfnamefont {U.}~\bibnamefont {Parts}},
    \bibinfo {author} {\bibfnamefont {P.~J.}\ \bibnamefont {Hakonen}}, \bibinfo
    {author} {\bibfnamefont {M.~A.}\ \bibnamefont {Paalanen}},\ and\ \bibinfo
    {author} {\bibfnamefont {E.~B.}\ \bibnamefont {Sonin}},\ }\bibfield  {title}
    {\bibinfo {title} {``superconductor-insulator transition'' in a single
    josephson junction},\ }\href {https://doi.org/10.1103/PhysRevLett.82.1004}
    {\bibfield  {journal} {\bibinfo  {journal} {Phys. Rev. Lett.}\ }\textbf
    {\bibinfo {volume} {82}},\ \bibinfo {pages} {1004} (\bibinfo {year}
    {1999})}\BibitemShut {NoStop}%
  \bibitem [{\citenamefont {Kuzmin}\ \emph {et~al.}(1991)\citenamefont {Kuzmin},
    \citenamefont {Nazarov}, \citenamefont {Haviland}, \citenamefont {Delsing},\
    and\ \citenamefont {Claeson}}]{kuzmin_coulomb_1991}%
    \BibitemOpen
    \bibfield  {author} {\bibinfo {author} {\bibfnamefont {L.~S.}\ \bibnamefont
    {Kuzmin}}, \bibinfo {author} {\bibfnamefont {Y.~V.}\ \bibnamefont {Nazarov}},
    \bibinfo {author} {\bibfnamefont {D.~B.}\ \bibnamefont {Haviland}}, \bibinfo
    {author} {\bibfnamefont {P.}~\bibnamefont {Delsing}},\ and\ \bibinfo {author}
    {\bibfnamefont {T.}~\bibnamefont {Claeson}},\ }\bibfield  {title} {\bibinfo
    {title} {Coulomb blockade and incoherent tunneling of cooper pairs in
    ultrasmall junctions affected by strong quantum fluctuations},\ }\href
    {https://doi.org/10.1103/PhysRevLett.67.1161} {\bibfield  {journal} {\bibinfo
     {journal} {Phys. Rev. Lett.}\ }\textbf {\bibinfo {volume} {67}},\ \bibinfo
    {pages} {1161} (\bibinfo {year} {1991})}\BibitemShut {NoStop}%
  \bibitem [{\citenamefont {Murani}\ \emph {et~al.}(2020)\citenamefont {Murani},
    \citenamefont {Bourlet}, \citenamefont {le~Sueur}, \citenamefont {Portier},
    \citenamefont {Altimiras}, \citenamefont {Esteve}, \citenamefont {Grabert},
    \citenamefont {Stockburger}, \citenamefont {Ankerhold},\ and\ \citenamefont
    {Joyez}}]{murani_absence_2020}%
    \BibitemOpen
    \bibfield  {author} {\bibinfo {author} {\bibfnamefont {A.}~\bibnamefont
    {Murani}}, \bibinfo {author} {\bibfnamefont {N.}~\bibnamefont {Bourlet}},
    \bibinfo {author} {\bibfnamefont {H.}~\bibnamefont {le~Sueur}}, \bibinfo
    {author} {\bibfnamefont {F.}~\bibnamefont {Portier}}, \bibinfo {author}
    {\bibfnamefont {C.}~\bibnamefont {Altimiras}}, \bibinfo {author}
    {\bibfnamefont {D.}~\bibnamefont {Esteve}}, \bibinfo {author} {\bibfnamefont
    {H.}~\bibnamefont {Grabert}}, \bibinfo {author} {\bibfnamefont
    {J.}~\bibnamefont {Stockburger}}, \bibinfo {author} {\bibfnamefont
    {J.}~\bibnamefont {Ankerhold}},\ and\ \bibinfo {author} {\bibfnamefont
    {P.}~\bibnamefont {Joyez}},\ }\bibfield  {title} {\bibinfo {title} {Absence
    of a dissipative quantum phase transition in josephson junctions},\ }\href
    {https://doi.org/10.1103/PhysRevX.10.021003} {\bibfield  {journal} {\bibinfo
    {journal} {Phys. Rev. X}\ }\textbf {\bibinfo {volume} {10}},\ \bibinfo
    {pages} {021003} (\bibinfo {year} {2020})}\BibitemShut {NoStop}%
  \bibitem [{\citenamefont {Masuki}\ \emph {et~al.}(2022)\citenamefont {Masuki},
    \citenamefont {Sudo}, \citenamefont {Oshikawa},\ and\ \citenamefont
    {Ashida}}]{masuki_absence_2022}%
    \BibitemOpen
    \bibfield  {author} {\bibinfo {author} {\bibfnamefont {K.}~\bibnamefont
    {Masuki}}, \bibinfo {author} {\bibfnamefont {H.}~\bibnamefont {Sudo}},
    \bibinfo {author} {\bibfnamefont {M.}~\bibnamefont {Oshikawa}},\ and\
    \bibinfo {author} {\bibfnamefont {Y.}~\bibnamefont {Ashida}},\ }\bibfield
    {title} {\bibinfo {title} {Absence versus presence of dissipative quantum
    phase transition in josephson junctions},\ }\href
    {https://doi.org/10.1103/PhysRevLett.129.087001} {\bibfield  {journal}
    {\bibinfo  {journal} {Phys. Rev. Lett.}\ }\textbf {\bibinfo {volume} {129}},\
    \bibinfo {pages} {087001} (\bibinfo {year} {2022})}\BibitemShut {NoStop}%
  \bibitem [{\citenamefont {Yang}\ \emph {et~al.}(2022)\citenamefont {Yang},
    \citenamefont {Liu}, \citenamefont {Liu}, \citenamefont {Wang}, \citenamefont
    {Qiu}, \citenamefont {Wang}, \citenamefont {Wang}, \citenamefont {He},
    \citenamefont {Li}, \citenamefont {Li}, \citenamefont {Tang}, \citenamefont
    {Wang}, \citenamefont {Xie}, \citenamefont {Valles}, \citenamefont {Xiong},\
    and\ \citenamefont {Li}}]{yang_signatures_2022}%
    \BibitemOpen
    \bibfield  {author} {\bibinfo {author} {\bibfnamefont {C.}~\bibnamefont
    {Yang}}, \bibinfo {author} {\bibfnamefont {H.}~\bibnamefont {Liu}}, \bibinfo
    {author} {\bibfnamefont {Y.}~\bibnamefont {Liu}}, \bibinfo {author}
    {\bibfnamefont {J.}~\bibnamefont {Wang}}, \bibinfo {author} {\bibfnamefont
    {D.}~\bibnamefont {Qiu}}, \bibinfo {author} {\bibfnamefont {S.}~\bibnamefont
    {Wang}}, \bibinfo {author} {\bibfnamefont {Y.}~\bibnamefont {Wang}}, \bibinfo
    {author} {\bibfnamefont {Q.}~\bibnamefont {He}}, \bibinfo {author}
    {\bibfnamefont {X.}~\bibnamefont {Li}}, \bibinfo {author} {\bibfnamefont
    {P.}~\bibnamefont {Li}}, \bibinfo {author} {\bibfnamefont {Y.}~\bibnamefont
    {Tang}}, \bibinfo {author} {\bibfnamefont {J.}~\bibnamefont {Wang}}, \bibinfo
    {author} {\bibfnamefont {X.~C.}\ \bibnamefont {Xie}}, \bibinfo {author}
    {\bibfnamefont {J.~M.}\ \bibnamefont {Valles}}, \bibinfo {author}
    {\bibfnamefont {J.}~\bibnamefont {Xiong}},\ and\ \bibinfo {author}
    {\bibfnamefont {Y.}~\bibnamefont {Li}},\ }\bibfield  {title} {\bibinfo
    {title} {Signatures of a strange metal in a bosonic system},\ }\href
    {https://doi.org/10.1038/s41586-021-04239-y} {\bibfield  {journal} {\bibinfo
    {journal} {Nature}\ }\textbf {\bibinfo {volume} {601}},\ \bibinfo {pages}
    {205} (\bibinfo {year} {2022})}\BibitemShut {NoStop}%
  \bibitem [{\citenamefont {Ambegaokar}\ and\ \citenamefont
    {Baratoff}(1963)}]{ambegaokar_tunneling_1963}%
    \BibitemOpen
    \bibfield  {author} {\bibinfo {author} {\bibfnamefont {V.}~\bibnamefont
    {Ambegaokar}}\ and\ \bibinfo {author} {\bibfnamefont {A.}~\bibnamefont
    {Baratoff}},\ }\bibfield  {title} {\bibinfo {title} {Tunneling between
    superconductors},\ }\href {https://doi.org/10.1103/PhysRevLett.10.486}
    {\bibfield  {journal} {\bibinfo  {journal} {Phys. Rev. Lett.}\ }\textbf
    {\bibinfo {volume} {10}},\ \bibinfo {pages} {486} (\bibinfo {year}
    {1963})}\BibitemShut {NoStop}%
  \bibitem [{\citenamefont {Kerr}\ \emph {et~al.}(1992)\citenamefont {Kerr},
    \citenamefont {Pan}, \citenamefont {Lichtenberger},\ and\ \citenamefont
    {Lea}}]{kerr_progress_1992}%
    \BibitemOpen
    \bibfield  {author} {\bibinfo {author} {\bibfnamefont {A.}~\bibnamefont
    {Kerr}}, \bibinfo {author} {\bibfnamefont {S.-K.}\ \bibnamefont {Pan}},
    \bibinfo {author} {\bibfnamefont {A.}~\bibnamefont {Lichtenberger}},\ and\
    \bibinfo {author} {\bibfnamefont {D.}~\bibnamefont {Lea}},\ }\bibfield
    {title} {\bibinfo {title} {Progress on tunerless sis mixers for the 200-300
    ghz band},\ }\href {https://doi.org/10.1109/75.165642} {\bibfield  {journal}
    {\bibinfo  {journal} {IEEE Microwave and Guided Wave Letters}\ }\textbf
    {\bibinfo {volume} {2}},\ \bibinfo {pages} {454} (\bibinfo {year}
    {1992})}\BibitemShut {NoStop}%
  \bibitem [{\citenamefont {Bruin}\ \emph {et~al.}(2013)\citenamefont {Bruin},
    \citenamefont {Sakai}, \citenamefont {Perry},\ and\ \citenamefont
    {Mackenzie}}]{bruin_similarity_2013}%
    \BibitemOpen
    \bibfield  {author} {\bibinfo {author} {\bibfnamefont {J.~A.~N.}\
    \bibnamefont {Bruin}}, \bibinfo {author} {\bibfnamefont {H.}~\bibnamefont
    {Sakai}}, \bibinfo {author} {\bibfnamefont {R.~S.}\ \bibnamefont {Perry}},\
    and\ \bibinfo {author} {\bibfnamefont {A.~P.}\ \bibnamefont {Mackenzie}},\
    }\bibfield  {title} {\bibinfo {title} {Similarity of scattering rates in
    metals showing ${T}$-linear resistivity},\ }\href
    {https://doi.org/10.1126/science.1227612} {\bibfield  {journal} {\bibinfo
    {journal} {Science}\ }\textbf {\bibinfo {volume} {339}},\ \bibinfo {pages}
    {804} (\bibinfo {year} {2013})}\BibitemShut {NoStop}%
  \bibitem [{\citenamefont {Legros}\ \emph {et~al.}(2019)\citenamefont {Legros},
    \citenamefont {Benhabib}, \citenamefont {Tabis}, \citenamefont
    {Lalibert{\'e}}, \citenamefont {Dion}, \citenamefont {Lizaire}, \citenamefont
    {Vignolle}, \citenamefont {Vignolles}, \citenamefont {Raffy}, \citenamefont
    {Li}, \citenamefont {Auban-Senzier}, \citenamefont {Doiron-Leyraud},
    \citenamefont {Fournier}, \citenamefont {Colson}, \citenamefont {Taillefer},\
    and\ \citenamefont {Proust}}]{legros_universal_2019}%
    \BibitemOpen
    \bibfield  {author} {\bibinfo {author} {\bibfnamefont {A.}~\bibnamefont
    {Legros}}, \bibinfo {author} {\bibfnamefont {S.}~\bibnamefont {Benhabib}},
    \bibinfo {author} {\bibfnamefont {W.}~\bibnamefont {Tabis}}, \bibinfo
    {author} {\bibfnamefont {F.}~\bibnamefont {Lalibert{\'e}}}, \bibinfo {author}
    {\bibfnamefont {M.}~\bibnamefont {Dion}}, \bibinfo {author} {\bibfnamefont
    {M.}~\bibnamefont {Lizaire}}, \bibinfo {author} {\bibfnamefont
    {B.}~\bibnamefont {Vignolle}}, \bibinfo {author} {\bibfnamefont
    {D.}~\bibnamefont {Vignolles}}, \bibinfo {author} {\bibfnamefont
    {H.}~\bibnamefont {Raffy}}, \bibinfo {author} {\bibfnamefont {Z.~Z.}\
    \bibnamefont {Li}}, \bibinfo {author} {\bibfnamefont {P.}~\bibnamefont
    {Auban-Senzier}}, \bibinfo {author} {\bibfnamefont {N.}~\bibnamefont
    {Doiron-Leyraud}}, \bibinfo {author} {\bibfnamefont {P.}~\bibnamefont
    {Fournier}}, \bibinfo {author} {\bibfnamefont {D.}~\bibnamefont {Colson}},
    \bibinfo {author} {\bibfnamefont {L.}~\bibnamefont {Taillefer}},\ and\
    \bibinfo {author} {\bibfnamefont {C.}~\bibnamefont {Proust}},\ }\bibfield
    {title} {\bibinfo {title} {Universal t-linear resistivity and planckian
    dissipation in overdoped cuprates},\ }\href
    {https://doi.org/10.1038/s41567-018-0334-2} {\bibfield  {journal} {\bibinfo
    {journal} {Nature Physics}\ }\textbf {\bibinfo {volume} {15}},\ \bibinfo
    {pages} {142} (\bibinfo {year} {2019})}\BibitemShut {NoStop}%
  \bibitem [{\citenamefont {Cao}\ \emph {et~al.}(2020)\citenamefont {Cao},
    \citenamefont {Chowdhury}, \citenamefont {Rodan-Legrain}, \citenamefont
    {Rubies-Bigorda}, \citenamefont {Watanabe}, \citenamefont {Taniguchi},
    \citenamefont {Senthil},\ and\ \citenamefont
    {Jarillo-Herrero}}]{cao_strange_2020}%
    \BibitemOpen
    \bibfield  {author} {\bibinfo {author} {\bibfnamefont {Y.}~\bibnamefont
    {Cao}}, \bibinfo {author} {\bibfnamefont {D.}~\bibnamefont {Chowdhury}},
    \bibinfo {author} {\bibfnamefont {D.}~\bibnamefont {Rodan-Legrain}}, \bibinfo
    {author} {\bibfnamefont {O.}~\bibnamefont {Rubies-Bigorda}}, \bibinfo
    {author} {\bibfnamefont {K.}~\bibnamefont {Watanabe}}, \bibinfo {author}
    {\bibfnamefont {T.}~\bibnamefont {Taniguchi}}, \bibinfo {author}
    {\bibfnamefont {T.}~\bibnamefont {Senthil}},\ and\ \bibinfo {author}
    {\bibfnamefont {P.}~\bibnamefont {Jarillo-Herrero}},\ }\bibfield  {title}
    {\bibinfo {title} {Strange metal in magic-angle graphene with near planckian
    dissipation},\ }\href {https://doi.org/10.1103/PhysRevLett.124.076801}
    {\bibfield  {journal} {\bibinfo  {journal} {Phys. Rev. Lett.}\ }\textbf
    {\bibinfo {volume} {124}},\ \bibinfo {pages} {076801} (\bibinfo {year}
    {2020})}\BibitemShut {NoStop}%
  \bibitem [{\citenamefont {Grissonnanche}\ \emph {et~al.}(2021)\citenamefont
    {Grissonnanche}, \citenamefont {Fang}, \citenamefont {Legros}, \citenamefont
    {Verret}, \citenamefont {Lalibert{\'e}}, \citenamefont {Collignon},
    \citenamefont {Zhou}, \citenamefont {Graf}, \citenamefont {Goddard},
    \citenamefont {Taillefer},\ and\ \citenamefont
    {Ramshaw}}]{grissonance_linear-in_2021}%
    \BibitemOpen
    \bibfield  {author} {\bibinfo {author} {\bibfnamefont {G.}~\bibnamefont
    {Grissonnanche}}, \bibinfo {author} {\bibfnamefont {Y.}~\bibnamefont {Fang}},
    \bibinfo {author} {\bibfnamefont {A.}~\bibnamefont {Legros}}, \bibinfo
    {author} {\bibfnamefont {S.}~\bibnamefont {Verret}}, \bibinfo {author}
    {\bibfnamefont {F.}~\bibnamefont {Lalibert{\'e}}}, \bibinfo {author}
    {\bibfnamefont {C.}~\bibnamefont {Collignon}}, \bibinfo {author}
    {\bibfnamefont {J.}~\bibnamefont {Zhou}}, \bibinfo {author} {\bibfnamefont
    {D.}~\bibnamefont {Graf}}, \bibinfo {author} {\bibfnamefont {P.~A.}\
    \bibnamefont {Goddard}}, \bibinfo {author} {\bibfnamefont {L.}~\bibnamefont
    {Taillefer}},\ and\ \bibinfo {author} {\bibfnamefont {B.~J.}\ \bibnamefont
    {Ramshaw}},\ }\bibfield  {title} {\bibinfo {title} {Linear-in temperature
    resistivity from an isotropic planckian scattering rate},\ }\href
    {https://doi.org/10.1038/s41586-021-03697-8} {\bibfield  {journal} {\bibinfo
    {journal} {Nature}\ }\textbf {\bibinfo {volume} {595}},\ \bibinfo {pages}
    {667} (\bibinfo {year} {2021})}\BibitemShut {NoStop}%
  \bibitem [{\citenamefont {Jaoui}\ \emph {et~al.}(2022)\citenamefont {Jaoui},
    \citenamefont {Das}, \citenamefont {Di~Battista}, \citenamefont
    {D{\'\i}ez-M{\'e}rida}, \citenamefont {Lu}, \citenamefont {Watanabe},
    \citenamefont {Taniguchi}, \citenamefont {Ishizuka}, \citenamefont
    {Levitov},\ and\ \citenamefont {Efetov}}]{jaoui_quantum_2022}%
    \BibitemOpen
    \bibfield  {author} {\bibinfo {author} {\bibfnamefont {A.}~\bibnamefont
    {Jaoui}}, \bibinfo {author} {\bibfnamefont {I.}~\bibnamefont {Das}}, \bibinfo
    {author} {\bibfnamefont {G.}~\bibnamefont {Di~Battista}}, \bibinfo {author}
    {\bibfnamefont {J.}~\bibnamefont {D{\'\i}ez-M{\'e}rida}}, \bibinfo {author}
    {\bibfnamefont {X.}~\bibnamefont {Lu}}, \bibinfo {author} {\bibfnamefont
    {K.}~\bibnamefont {Watanabe}}, \bibinfo {author} {\bibfnamefont
    {T.}~\bibnamefont {Taniguchi}}, \bibinfo {author} {\bibfnamefont
    {H.}~\bibnamefont {Ishizuka}}, \bibinfo {author} {\bibfnamefont
    {L.}~\bibnamefont {Levitov}},\ and\ \bibinfo {author} {\bibfnamefont {D.~K.}\
    \bibnamefont {Efetov}},\ }\bibfield  {title} {\bibinfo {title} {Quantum
    critical behaviour in magic-angle twisted bilayer graphene},\ }\href
    {https://doi.org/10.1038/s41567-022-01556-5} {\bibfield  {journal} {\bibinfo
    {journal} {Nature Physics}\ }\textbf {\bibinfo {volume} {18}},\ \bibinfo
    {pages} {633} (\bibinfo {year} {2022})}\BibitemShut {NoStop}%
  \bibitem [{\citenamefont {Pekola}\ \emph {et~al.}(1994)\citenamefont {Pekola},
    \citenamefont {Hirvi}, \citenamefont {Kauppinen},\ and\ \citenamefont
    {Paalanen}}]{pekola_thermometry_1994}%
    \BibitemOpen
    \bibfield  {author} {\bibinfo {author} {\bibfnamefont {J.~P.}\ \bibnamefont
    {Pekola}}, \bibinfo {author} {\bibfnamefont {K.~P.}\ \bibnamefont {Hirvi}},
    \bibinfo {author} {\bibfnamefont {J.~P.}\ \bibnamefont {Kauppinen}},\ and\
    \bibinfo {author} {\bibfnamefont {M.~A.}\ \bibnamefont {Paalanen}},\
    }\bibfield  {title} {\bibinfo {title} {Thermometry by arrays of tunnel
    junctions},\ }\href {https://doi.org/10.1103/PhysRevLett.73.2903} {\bibfield
    {journal} {\bibinfo  {journal} {Phys. Rev. Lett.}\ }\textbf {\bibinfo
    {volume} {73}},\ \bibinfo {pages} {2903} (\bibinfo {year}
    {1994})}\BibitemShut {NoStop}%
  \bibitem [{\citenamefont {Hirvi}\ \emph {et~al.}(1995)\citenamefont {Hirvi},
    \citenamefont {Kauppinen}, \citenamefont {Korotkov}, \citenamefont
    {Paalanen},\ and\ \citenamefont {Pekola}}]{hirvi_arrays_1995}%
    \BibitemOpen
    \bibfield  {author} {\bibinfo {author} {\bibfnamefont {K.~P.}\ \bibnamefont
    {Hirvi}}, \bibinfo {author} {\bibfnamefont {J.~P.}\ \bibnamefont
    {Kauppinen}}, \bibinfo {author} {\bibfnamefont {A.~N.}\ \bibnamefont
    {Korotkov}}, \bibinfo {author} {\bibfnamefont {M.~A.}\ \bibnamefont
    {Paalanen}},\ and\ \bibinfo {author} {\bibfnamefont {J.~P.}\ \bibnamefont
    {Pekola}},\ }\bibfield  {title} {\bibinfo {title} {Arrays of normal metal
    tunnel junctions in weak coulomb blockade regime},\ }\href
    {https://doi.org/10.1063/1.115090} {\bibfield  {journal} {\bibinfo  {journal}
    {Applied Physics Letters}\ }\textbf {\bibinfo {volume} {67}},\ \bibinfo
    {pages} {2096} (\bibinfo {year} {1995})}\BibitemShut {NoStop}%
  \end{thebibliography}

\section*{Acknowledgments}
We thank David Haviland, Jukka Pekola, Anton Bubis and Alexander Shnirman for helpful feedback on the manuscript.
This research was supported by the Scientific Service Units of IST Austria through resources provided by the MIBA Machine Shop and the Nanofabrication Facility.
Work funded by Austrian FWF grant P33692-N.
J.S. acknowledges funding from the European Union's Horizon 2020 research and innovation program under the Marie Skłodowska-Curie Grant Agreement No. 754411.

\section*{Supplementary materials}
Raw data for all figures to be uploaded before final publication\\
Materials and Methods\\
Supplementary Text\\
Figs. S1 to S15\\
References \cite{cedergren_insulating_2017,ambegaokar_tunneling_1963,kerr_progress_1992,bruin_similarity_2013,yang_signatures_2022,legros_universal_2019,cao_strange_2020,grissonance_linear-in_2021,jaoui_quantum_2022,pekola_thermometry_1994,hirvi_arrays_1995,bard_superconductor-insulator_2017,sondhi_continuous_1997,giamarchi_anderson_1988}

\newcommand{\suppage}[1]{
	\pagebreak
	\begin{figure}[p]
		\vspace*{-1.5cm}
		\hspace*{-1.9cm}
		\includegraphics[page=#1]{supplement.pdf}
		\centering
	\end{figure}
}

\suppage{1}
\suppage{2}
\suppage{3}
\suppage{4}
\suppage{5}
\suppage{6}
\suppage{7}
\suppage{8}
\suppage{9}
\suppage{10}
\suppage{11}
\suppage{12}
\suppage{13}
\suppage{14}
\suppage{15}
\suppage{16}
\suppage{17}
\suppage{18}
 
\end{document}